\documentclass{llncs}
\usepackage{amsmath}
\usepackage{svg}
\usepackage{booktabs}
\usepackage{graphicx}
\usepackage[flushleft]{threeparttable}
\usepackage[symbol]{footmisc}

\begin{document}
	\title{An Open Framework Enabling Electromagnetic Tracking in Image-Guided Interventions}
	\titlerunning{}
	\toctitle{}
	
	\author{}
	\institute{}
	\author{Herman Alexander Jaeger\inst{1}, Stephen Hinds\inst{1} \and P\'adraig Cantillon-Murphy\inst{1}\inst{2}\inst{3}}
	\institute{University College Cork, Cork, Ireland
\and Tyndall National Institute, Cork, Ireland
\and Institute of Image Guided Surgery, Strasbourg, France}
	
	\maketitle

	\begin{abstract}
	Electromagnetic tracking (EMT) is a core platform technology in the navigation and visualisation of image-guided procedures. The technology provides high tracking accuracy in non-line-of-sight environments, allowing instrument navigation in locations where optical tracking is not feasible. Integration of EMT in complex procedures, often coupled with multi-modal imaging, is on the rise, yet the lack of flexibility in the available hardware platforms has been noted by many researchers and system designers. Advances in the field of EMT include novel methods of improving tracking system accuracy, precision and error compensation capabilities, though such system-level improvements cannot be readily incorporated in current therapy applications due to the `blackbox' nature of commercial tracking solving algorithms. This paper defines a software framework to allow novel EMT designs and improvements become part of the global design process for image-guided interventions. In an effort to standardise EMT development, we define a generalised cross-platform software framework in terms of the four system functions common to all EMT systems; \textit{acquisition},  \textit{filtering},  \textit{modelling} and  \textit{solving}. The interfaces between each software component are defined in terms of their input and output data structures. An exemplary framework is implemented in the Python programming language and demonstrated with the open-source Anser EMT system. Performance metrics are gathered from both Matlab and Python implementations of Anser EMT considering the host operating system, hardware configuration and acquisition settings used. Results show indicative system latencies of 5 ms can be achieved using the framework on a Windows operating system, with decreased system performance observed on UNIX-like platforms.
	\end{abstract}

	\section{Introduction}
\begin{figure}
		\centering
		\includegraphics[width=0.7\linewidth]{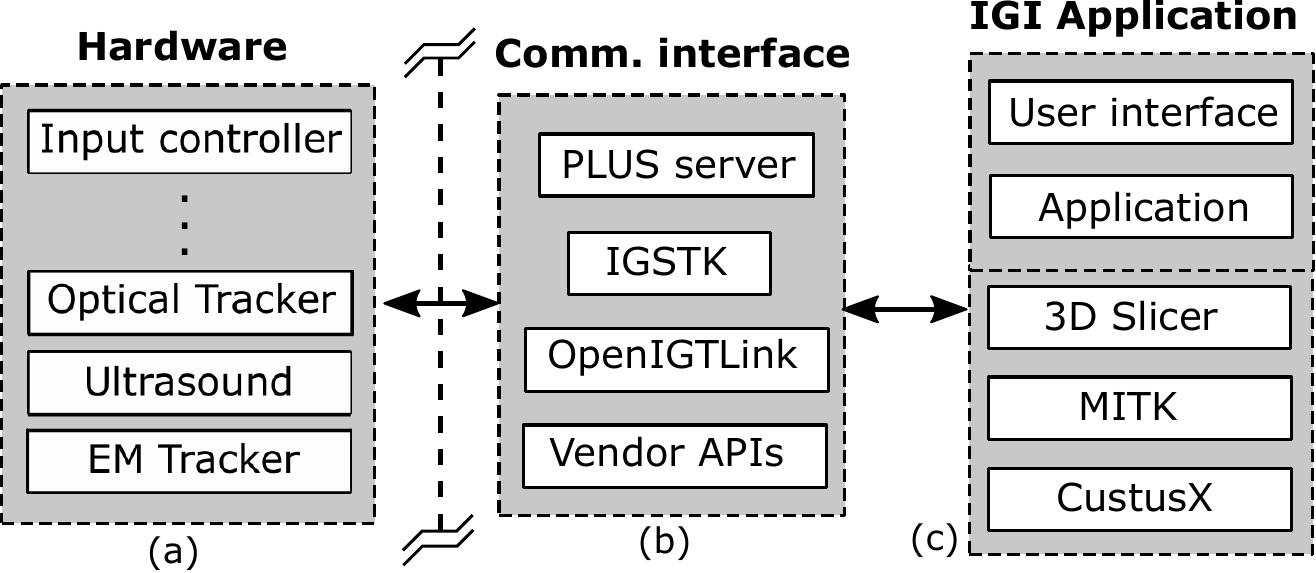}
		\caption{A standard approach for developing cross-platform IGI applications. (a) Proprietary hardware is interfaced using an API or open communications interface (b). (c) Guided therapy applications use this interface to ensure cross-platform compatibility.}
		\label{fig:workflow}
	\end{figure}

	The development of new image guided therapies relies heavily on intelligently combining data from multiple hardware sources. New techniques combining ultrasound and electromagnetic tracking (EMT) \cite{Franz2017,Paolucci2018} are among techniques which combine multiple data sources to enhance the safety and accuracy of procedures. Progress in these areas is made possible by the standardised open protocols \cite{Lasso2014,Tokuda2009} that govern how hardware and software should interact with one another. Fig. \ref{fig:workflow} shows the generalised design flow of many image guided interventions (IGI). IGI applications interact with hardware through vendor authored application programming interfaces (APIs) or interface toolkits such as PLUS and IGSTK\cite{Enquobahrie2007}. Such toolkits provide standardised methods through which IGI applications and hardware can interact. From a software perspective, this standardised approach enables developers to prototype and apply their work in a manner that can be distributed and replicated. That said, IGI application development typically falls short of incorporating custom innovations in tracking hardware.  Electromagnetic tracking systems in particular are very much considered `blackboxes' from the perspective of IGI research.

\begin{figure}
 	\centering
 	\includegraphics[width=0.45\linewidth]{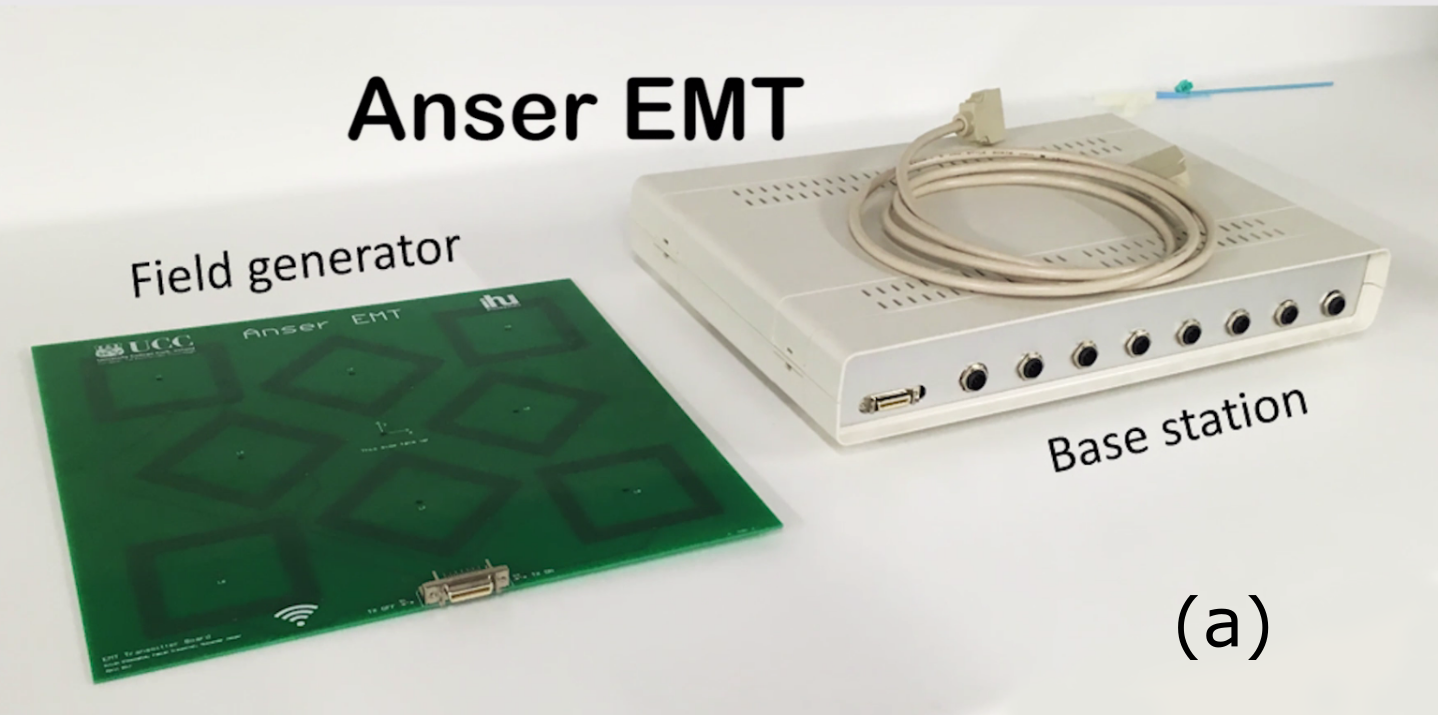}
 	\includegraphics[width=0.35\linewidth]{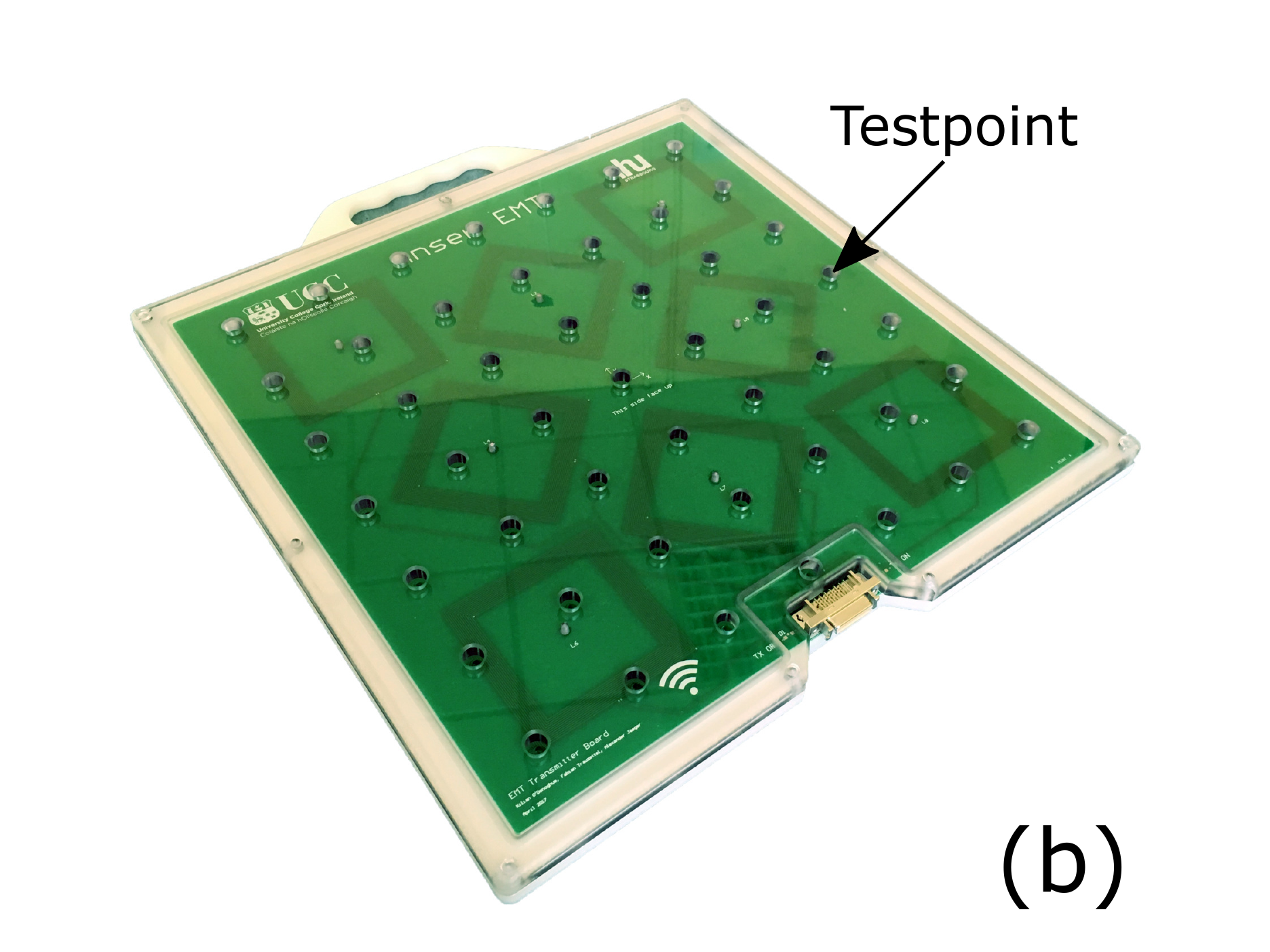}
 	\caption{(a) The Anser EMT system v1.0. (b) Field generator enclosure.}
 	\label{fig:anser}
 	\end{figure}

 	This paper outlines a framework to encourage integration of new electromagnetic tracking hardware into the current IGI design flow. The resulting framework was implemented in the Python programming language and applied to the open-source Anser EMT system \cite{Jaeger2017} shown in Fig. \ref{fig:anser}. Preliminary cross-platform functionality of the framework is demonstrated with important performance metrics reported.

	\section{Framework Design}
\begin{figure}
		\centering
		\includegraphics[width=0.7\linewidth]{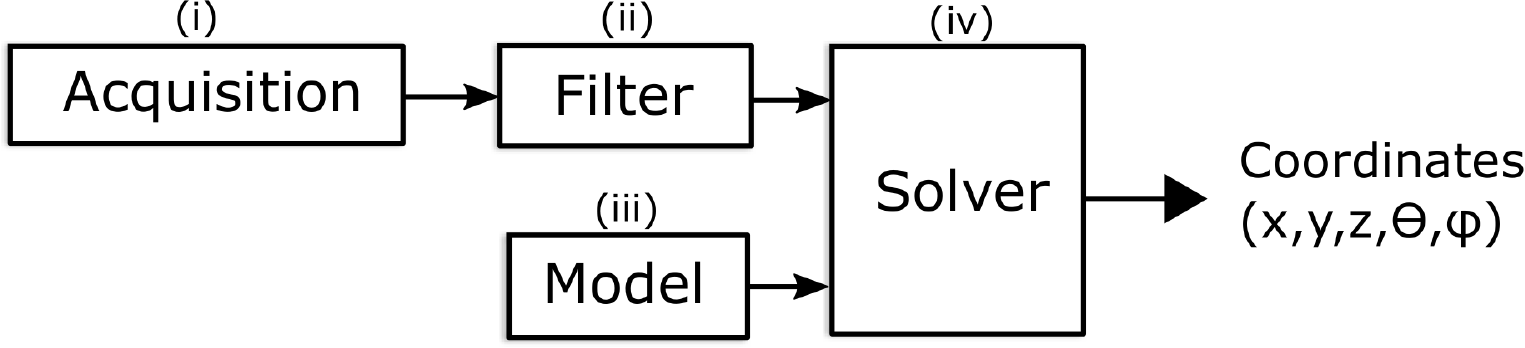}
		\caption{The basic components which comprise all electromagnetic tracking systems are shown. The 5 degree-of-freedom sensor coordinate $[x,y,z,\theta,\varphi]$ is the most basic case for a symmetrical sensor coil, where $\theta$ and $\varphi$ correspond to yaw and pitch angles respectively.}
		\label{fig:emtdesign}
	\end{figure}

	Electromagnetic tracking systems are complex electronic systems that incorporate advanced analog circuit design, signal processing and optimisation techniques. While the precise topology of such systems will vary depending on the core design and manufacturer, all EMT systems can be distinguished by four core processing steps outlined in Fig. \ref{fig:emtdesign}: \textit{acquisition}, \textit{filtering}, \textit{modelling}, and \textit{solving}. Each of these processing steps can be treated as a discrete, independent stage in the tracking system software pipeline.  The designed framework is structured according to these stages.

	\subsubsection{(i) Acquisition.} The most fundamental step in any measurement system is data acquisition. It is the process by which physical signals are digitised into forms suitable for computation. In the case of EMT systems, the signals being measured are typically electric voltages. Such signals are induced on a tracking sensor coil when it is placed in the tracking volume of an EMT system. Acquisition hardware in the form of an analogue-to-digital converter (ADC) converts the electrical signal into a digital data stream by sampling at fixed periodic intervals. Manufacturers of acquisition systems include National Instruments (Austin, TX, U.S.A.) and Measurement Computing Corp. (Norton, MA, U.S.A).
	\subsubsection{(ii) Filtering.} The sampled sensor signal contains all the necessary physical information required to resolve the sensor's position in space. EMT systems typically operate using multiple transmitters operating at distinct frequencies, thus the sampled sensor signal is a linear sum of the individual frequency components generated by each transmitting coil, as well as noise from the surrounding environment. The filter extracts the relevant signal content. A combination of digital filtering and Fourier methods are typically employed.
	\subsubsection{(iii) Model.} An accurate model of the tracking system's generated magnetic field is a necessary component in all system designs. Models are typically defined as analytical expressions which define the spacial distribution of magnetic fields in the volume around the field transmitter. Each constituent transmitting coil can be characterised by a vector equation relating the magnetic flux density vector $\vec{B}$ to a point in space, $\vec{p}$:
	\begin{equation}
	\vec{B(\vec{p})} = [B_x^p,B_y^p,B_z^p]
	\label{eq:model}
	\end{equation}
	Commonly used models include variations of the magnetic dipole approximation \cite{Li2013a}, Biot-Savart law \cite{Kilian2014-Tracking,Sonntag2007} and mutual inductance models \cite{Bien2014}. Numerical models may also be used in cases where no accurate closed-form solution for the magnetic field exists.

	\subsubsection{(iv) Solver.} Magnetic tracking systems resolve sensor positions through a process of non-linear optimisation in which a cost function is minimised to yield the best-fit solution for the position and orientation of the tracking sensor. The cost function is generally formulated such that the squared difference between the magnetic model of the sensor coil and acquired sensor measurements are minimised. Full formulations of the sensor model can be found in \cite{Kilian2014-Tracking}. A non-linear least-squares approach is usually required to yield accurate results. Examples of general solving methods include the well known Levenburg-Marquard \cite{Levenberg1944} and trust-region algorithms \cite{Byrd1987}. The general form of the optimisation problem is shown in \eqref{eq:optimisationproblem}:	
	\begin{equation}
	\begin{aligned}
	& \underset{\mathbf{p}}{\text{minimise}}
	& & \sum f_i(\mathbf{p})^2 \\
	& \text{subject to}
	& & \textit{lb} \leq \mathbf{p} \leq \textit{ub}, i = 1, \ldots, n \\
	\end{aligned}
	\label{eq:optimisationproblem}
	\end{equation}
	where $f_i$ is the $i^{th}$ cost function relating a single frequency component of the tracking sensor signal to the field model of the corresponding transmission coil, \textit{lb} and \textit{ub} are upper and lower bound constraints for the solving algorithm (if applicable) and $\mathbf{p}$ is the vector argument representing the position and orientation of the tracking sensor.

	\section{Framework Implementation}
	
\begin{figure}
	\centering
	\includegraphics[width=0.9\linewidth]{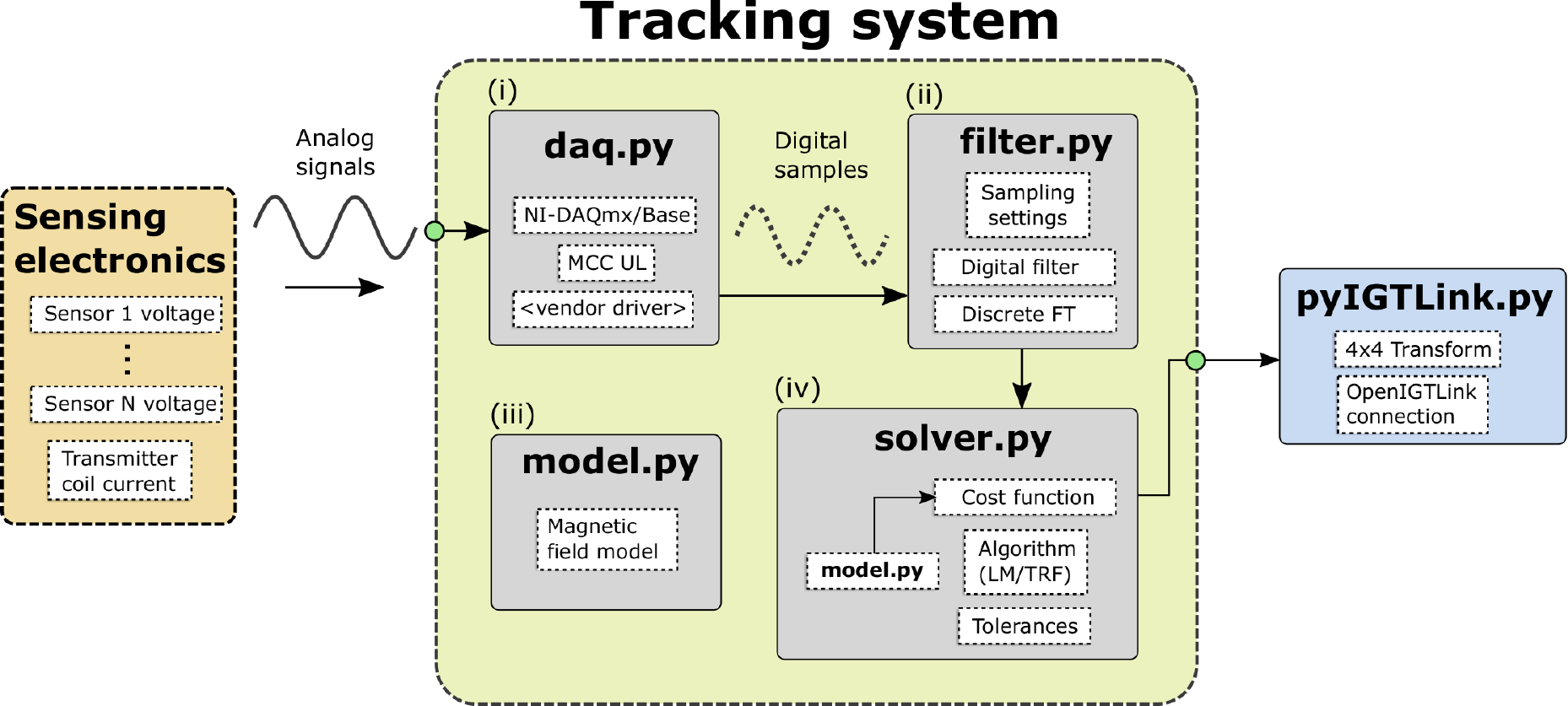}
	\caption{(i) Analogue signals are sampled by the data acquisition module. (ii) The samples are filtered in software with relevant frequency components extracted. (iii)-(iv) A cost function utilises the component magnitudes to yield a position vector which can be transmitted using OpenIGTLink as a 4x4 transformation matrix.}
	\label{fig:modules}
	\end{figure}

	The proposed EMT framework is composed of four Python modules representing each of the signal processing steps shown in Fig. \ref{fig:workflow}. An expansion of this design showing the data-flow between modules is shown in  Fig. \ref{fig:modules}. The framework is divided into four modules labelled (i) to (iv). Analog signals from the EMT sensing electronics are acquired through a data acquisition module \textit{daq.py}. This module provides facilities to abstract the acquisition hardware's specific API into a standard interface. The acquired digital samples are fed into the filter module \textit{filter.py} where the frequency components of interest are conditioned and extracted from the digital waveform. The filter module allows easy configuration of filter parameters while providing routines for efficient matrix multiplications required during filtering operations. The extracted signal information is then fed to the solver module \textit{solver.py}. Simultaneously the magnetic model of the system \textit{model.py} is compared with the extracted signal data to minimise the system cost function. The solver module provides access to the tolerance and parameter settings for the minimisation process. The resulting sensor coordinates from the solver can be streamed to the user application using OpenIGTLink \cite{Tokuda2010}.

	The proposed framework was applied to the Anser EMT project \cite{Jaeger2017}. The original open-source codebase was implemented in Matlab\footnote{Available at \url{https://openemt.org}} and was fully refactored to conform with the framework. The PyDAQmx driver \cite{Clade} was used to provide a cross-platform interface with the NI-DAQmx driver acquisition system (National Instruments, Austin, Texas). Filtering and solving operations were performed using NumPy \cite{VanDerWalt2011} and SciPy libraries\cite{Oliphant2006}. OpenIGTLink connectivity was achieved using PyIGTLink \cite{Hiversen2016}.
	
	\section{Experiments and Results}
	
	EMT framework's performance and accuracy were tested relative to the original Windows-only Matlab implementation. Testing of the framework was performed on three operating systems: Windows 10, MacOS 10.13 and Cent-OS 7.0 Linux. All reported metrics result from tests performed on a laptop PC configured with an Intel i7 4810HQ 3.5GHz CPU and 16GB of RAM, utilising both a Windows 10 and CentOS 7.0 installation. Compatibility with the MacOS 10.13 operating system was confirmed using a separate machine, but comparative results were not possible due to significant differences in the laptop hardware configuration. The variable sampling frequency of the acquisition system was set at 100kHz for all experiments.
	
	\subsection{Performance benchmark} 
	Performance testing measured the framework's ability to stream position measurements as quickly as possible with minimal latency over an OpenIGTLink connection to CustusX and 3DSlicer. Acquisition latency, maximum update frequency (with both a stationary and moving sensor) were recorded for multiple acquisition frame sizes shown in Table \ref{tab:pyperf}. An acquisition frame constitutes the number of samples gathered by the acquisition system per single resolved position. A finite acquisition time for each frame puts a limit on the minimum latency figure. Static refers to a slow moving sensor speeds of $<$5cm per second while `Dynamic' refers to speeds $>$50cm per second.
	\begin{table}[]	
		\centering
		\caption{Performance measurements in Matlab (Windows only) and Python (Windows/Linux). }
		\label{tab:matperf}
		\scalebox{0.78}{
			\begin{tabular}{@{}c|c|c|c|c@{}}
				\toprule
				\textbf{Matlab}&Frame size& Acq. Latency (ms)  &Max. Static (Hz)  &Max. Dynamic (Hz)   \\ \midrule
				-&250&2.5&  72&64 \\
				-&500&5&  70&61   \\
				-&1000&10& 65&62  \\
				-&2000&20&  50&50 \\
				-&5000&50&  20&20 \\ \bottomrule
			\end{tabular}
		}
			\centering
			\scalebox{0.78}{
			\begin{threeparttable}
			\label{tab:pyperf}
			\begin{tabular}{@{}c|c|c|c|c@{}}
				\toprule
				\textbf{Python}&Frame size&Acq. Latency (ms)  &Max. Static (Hz)  &Max. Dynamic (Hz)   \\ \midrule
				-&250&2.5/5000+* &138/120  &84/70     \\
				-&500&5/2000+&142/115  &95/83    \\
				-&1000&10/1000+&102/80  &66/62  \\
				-&2000&20/100&51/45  &52/45    \\
				-&5000&50/50&20/21  &20/20     \\ \bottomrule
			\end{tabular}
			\begin{tablenotes}
				\small
				\item * Due to operating system driver, see discussion.
			\end{tablenotes}
		\end{threeparttable}
	}
	\end{table}

	\subsection{Accuracy benchmark}
	Accuracy testing consisted of comparing errors between sensor positions obtained from a 7x7 plane test $x$-$y$ grid providing a total of 49 points. 150 position acquisitions at a height of $z$ = 70mm from the transmitter board (Fig. \ref{fig:anser} (b)) were recorded per grid point from which the mean $x$, $y$ and $z$ coordinate of each point was calculated. Measurements were obtained from both Matlab and Python implementations. Maximum, minimum and root-mean-square (RMS) errors were calculated between the two obtained point grids. The grid obtained using the Matlab implementation was used as the reference since its performance has already been characterised in \cite{Jaeger2017}. Maximum and minimum grid errors were measured as 3.1mm and 0.1mm respectively with an RMS error of 0.9mm with a standard deviation of 0.75mm.

	\section{Discussion}
	Table \ref{tab:pyperf} shows how performance of the Anser EMT system varies with acquisition sample size over Windows and Linux operating systems. Benchmark results on Windows are clearly favourable to Linux particularly at low sample sizes. The high latencies in Linux were found to be caused by a low-level buffer issue due to limitations in the NI-DAQmx Base kernel driver for Linux. Forcing the acquisition time to be greater than the solving time prevents this latency issue occurring. Empirically setting the acquisition frame to 2000 (i.e. a frame acquisition time of 20ms) mitigates the issue, although this limits the maximum effective position update rate.
	
	It can also be seen that the update speeds vary significantly depending on the movement speed of the sensor. This is due to the previous sensor position being used as the initial condition for the solver during each position update. A moving sensor causes previous sensor positions to lie further from the current true sensor position, resulting in the solver requiring more iterations to converge to the global minimum. Artificially forcing periodicity by limiting the update rate of the system prevents this issue from occurring. This approach, common in commercial platforms results in significantly increased system latency.
	
	The purpose of the accuracy benchmark is to showcase any significant differences between the two EMT software implementations. The benchmark is limited since it uses the Matlab implementation results as the reference standard, since no gold standard was available for the experiment. The reported mean error value of 0.9mm falls within the mean error value of the Matlab implementation of 1.14mm \cite{Jaeger2017}. From this we can conclude that the Python implementation is of similar accuracy to the original implementation. Characterisation of this system according to the Hummel protocol \cite{Hummel2006,Franz2014} is necessary in order to fully validate the accuracy of the system under the new framework.
	
	\subsection*{Conclusion}
	 An open source framework for designing electromagnetic tracking systems has been proposed. The framework was applied to a previously characterised tracking system with performance results reported. It is hoped that this work will assist in the translation of new EMT modules and platforms from research into the clinical setting.
	
	\bibliographystyle{splncs} 
	\bibliography{library}
	
\end{document}